\documentclass[letterpaper,12pt]{article}
\usepackage{amsmath}
\usepackage{amsfonts}
\usepackage{amssymb}
\usepackage{graphicx}
\usepackage{physics}
\usepackage{latexsym}
\usepackage{epsfig}
\usepackage{pstricks}
\usepackage{stmaryrd}
\usepackage{rotating}
\usepackage[english]{babel}
\usepackage{setspace}
\usepackage[utf8]{inputenc}
\usepackage[T1]{fontenc}
\usepackage{lmodern}
\usepackage{enumitem}
\usepackage{csquotes}
\usepackage{amsthm}
\usepackage{xcolor}
 
\begin{document}
\begin{center}	
\begin{LARGE}
\textbf{Wigner's convoluted friends}\\
\end{LARGE}
\end{center}

\begin{center}
\begin{large}
R. Muciño and E. Okon\\
\end{large}
\textit{Universidad Nacional Aut\'onoma de M\'exico, Mexico City, Mexico.}\\
\end{center}

Considering a complicated extension of a Wigner's friend scenario, Frauchiger and Renner (FR) allegedly showed that ``quantum theory cannot consistently describe the use of itself''. However, such a result has been under severe criticism, as it has been convincingly argued to crucially depend on an implicit, non-trivial assumption regarding details of the collapse mechanism. In consequence, the result is not as robust or general as intended. On top of all this, in this work we show that a much simpler arrangement---basically an EPR setting---is sufficient to derive a result fully analogous to that of FR. Moreover, we claim that all lessons learned from FR's result are essentially contained within the original EPR paper. We conclude that FR's result does not offer any novel insights into the conceptual problems of quantum theory.

\onehalfspacing
\section{Introduction}

In \cite{FR}, Frauchiger and Renner (FR) present a result allegedly proving that ``quantum theory cannot consistently describe the use of itself''. FR consider a complicated extension of a Wigner's friend scenario, involving two labs and four observers, and the conflict is argued to primarily arise from the assumption that quantum mechanics can be consistently applied to complex, macroscopic systems. In more detail, the result is cast in the form of a no-go theorem arguing for the mutual incompatibility of three assumptions: (C), demanding consistency between different observers, (S), establishing that measurements have single outcomes and (Q), (allegedly) capturing the universal validity of quantum mechanics.

However, as it has been convincingly argued elsewhere, \cite{baumann,sudbery1,sudbery2,lazarovici,Drezet,tausk}, the FR result seems to depend on an implicit, non-trivial, idiosyncratic interpretation of the quantum formalism, particularly regarding the collapse postulate. Concretely, in effect it is assumed that when a measurement is carried out inside of a closed lab, such a measurement leads to a collapse for inside observers, but does not do so for outside observers. Once one recognizes such an assumption, it comes as no surprise the fact that a contradiction is reached. Moreover, once one recognizes such an assumption, it comes as a surprise that an arrangement as complicated as that considered by FR is necessary to derive a conflict.

In this work we show that a much simpler setting, basically an EPR scenario, is sufficient to arrive at a result fully analogous to that of FR. Moreover, we claim that the true consequences of FR's result seem to contain not much more than what can be learned from the original EPR paper \cite{EPR}. With these objectives in mind, our short manuscript is organized as follows. In section \ref{FR} we describe FR's work and in section \ref{EPR} we show how a fully analogous result can be achieved in a much simpler setting: an EPR scenario. We also comment on what this means regarding potential consequences of FR's result. We close with some final comments in section \ref{C}.

\section{The FR argument}
\label{FR}

We start by describing the experimental arrangement considered by FR. It contains four different agents, F, $\overline{\text{F}}$, W and $\overline{\text{W}}$ and two labs, L and $\overline{\text{L}}$. F is inside L and $\overline{\text{F}}$ inside $\overline{\text{L}}$ and W and $\overline{\text{W}}$ are outside of the respective labs and can perform measurements on them. Finally, there is a communication channel from $\overline{\text{L}}$ to L.
 
The experiment runs in steps as follows:
\begin{description}[font=\normalfont,labelindent=.5cm]
\item[\textbf{Step 1}:]  $\overline{\text{F}}$ prepares a quantum coin in the state $\sqrt{\frac{1}{3}}\ket{h}+ \sqrt{\frac{2}{3}}\ket{t}$ and measures it. If she finds \emph{h}, she prepares an electron in the state $\ket{\downarrow}$, if she finds \emph{t}, she prepares it in the state $\ket{\rightarrow}= \frac{1}{\sqrt{2}} \left( \ket{\uparrow}+\ket{\downarrow} \right)$. She then sends the prepared electron to F.
\item[\textbf{Step 2}:] F measures the electron in the $\{\ket{\uparrow},\ket{\downarrow}\}$ basis.
\item[\textbf{Step 3}:] $\overline{\text{W}}$ measures $\overline{\text{L}}$ in the following basis:
$$
\ket{o}_{\overline{\text{L}}}=\frac{1}{\sqrt{2}}\left[\ket{h}_{\overline{\text{L}}}-\ket{t}_{\overline{\text{L}}}\right],
$$
$$
\ket{f}_{\overline{\text{L}}}=\frac{1}{\sqrt{2}}\left[\ket{h}_{\overline{\text{L}}}+\ket{t}_{\overline{\text{L}}}\right]
$$
(with $\ket{h}_{\overline{\text{L}}}$ and $\ket{t}_{\overline{\text{L}}}$ the states of $\overline{\text{L}}$ after $\overline{\text{F}}$ measures the coin and finds the corresponding result). $\overline{\text{W}}$ then announces her result.
\item[\textbf{Step 4}:] W measures L in the following basis:
$$
\ket{o}_{\text{L}}=\frac{1}{\sqrt{2}}\left[\ket{\downarrow}_{\text{L}}-\ket{\uparrow}_{\text{L}}\right],
$$
$$
\ket{f}_{\text{L}}=\frac{1}{\sqrt{2}}\left[\ket{\downarrow}_{\text{L}}+\ket{\uparrow}_{\text{L}}\right]
$$
(with $\ket{\downarrow}_{\text{L}}$ and $\ket{\uparrow}_{\text{L}}$ the states of L after F measures the spin and finds the corresponding result).
\end{description}

FR then analyze the experiment assuming (C), (S) and (Q), together with the non-trivial assumption mentioned in the introduction, namely:
\begin{description}[font=\normalfont,labelindent=.5cm]
\item[(H)] When a measurement is carried out inside of a closed lab, such a measurement leads to a collapse for inside observers, but it does not lead to a collapse for outside observers.
\end{description}
By doing so, they consider several implications (see Table 3 in \cite{FR}):
\begin{enumerate}
\item If $\overline{\text{F}}$ finds \emph{t}, then she sends F the electron in state $\ket{\rightarrow}$. If so, $\overline{\text{F}}$ reasons employing (H), after F measures, W will assign to L the state $\ket{f}_{\overline{\text{L}}}$ so W's measurement will result for sure in \emph{f}. That is:
\begin{equation}\label{I1}
\text{If $\overline{\text{F}}$ finds \emph{t}, she knows that W will find \emph{f} with certainty.}
\end{equation}
\item If F finds $\uparrow$, then she reasons that $\overline{\text{F}}$ must have gotten \emph{t}. But we know from (\ref{I1}) that that implies that W will find \emph{f} with certainty. Then, using (C) we conclude:
\begin{equation}\label{I2}
\text{If F finds $\uparrow$, then she knows that W will find \emph{f} with certainty.}
\end{equation}
\item Given (H), according to $\overline{\text{W}}$, after step 2 the system $\overline{\text{L}}$+L is in the state
$$
\sqrt{\frac{1}{3}}\ket{h}_{\overline{\text{L}}}\ket{\downarrow}_{\text{L}}+ \sqrt{\frac{2}{3}}\ket{t}_{\overline{\text{L}}}\ket{\rightarrow}_{\text{L}} .
$$
But that state is orthogonal to $\ket{o}_{\overline{\text{L}}}\ket{\downarrow}_{\text{L}}$. As a result, employing (\ref{I2}) and (C):
\begin{equation}\label{I3}
\text{If $\overline{\text{W}}$ finds \emph{o}, then she knows that W will find \emph{f} with certainty.}
\end{equation}
\item Finally, using (\ref{I3}) and (C) we conclude:
\begin{equation}\label{I4}
\text{If $\overline{\text{W}}$ announces she got \emph{o}, then W knows she will find \emph{f} with certainty.}
\end{equation}
\end{enumerate}

The problem is that, at the same time, given (H), according to W the state of L+$\overline{\text{L}}$ after step 2 is given by
$$
\sqrt{\frac{1}{3}}\ket{h}_{\overline{\text{L}}}\ket{\downarrow}_{\text{L}}+ \sqrt{\frac{2}{3}}\ket{t}_{\overline{\text{L}}}\ket{\rightarrow}_{\text{L}} ,
$$
which can also be written as
$$
 \frac{1}{2\sqrt{3}} \left[ \ket{o}_{\overline{\text{L}}}\ket{o}_{\text{L}} - \ket{o}_{\overline{\text{L}}} \ket{f}_{\text{L}}+\ket{f}_{\overline{\text{L}}} \ket{o}_{\text{L}} \right] + \frac{\sqrt{3}}{2}\ket{f}_{\overline{\text{L}}} \ket{f}_{\text{L}}.
$$
Given that $\ket{o}_{\overline{\text{L}}}\ket{o}_{\text{L}}$ has a non-zero coefficient, W concludes that it is possible for her to get the result \emph{o}, even if $W$ also gets \emph{o}. But this is inconsistent with (\ref{I4}).

As advertised, the conjunction of (S), (C), (Q) and (H) is inconsistent. To see that H indeed plays a crucial role in the argument, we note that, when $\overline{\text{F}}$ measures the coin, it is assumed that, even though the initial state is a superposition of \emph{h} and \emph{t}, she will find either \emph{h} or \emph{t} (and similarly for all other measurements). Moreover, for future predictions, $\overline{\text{F}}$ assumes her measurement caused a collapse. That is, for future predictions, $\overline{\text{F}}$ uses the state corresponding to what she found, not the full superposition. Only if this is the case, $\overline{\text{F}}$ will conclude that if she found \emph{t}, then W will find \emph{f} (see the first implication above). At the same time, $\overline{\text{W}}$, for instance, assigns to $\overline{\text{L}}$ the fully superposition of \emph{h} and \emph{t}, even after $\overline{\text{F}}$'s measurement. Only because of this, $\overline{\text{W}}$ concludes that if she finds \emph{o}, then she knows that W will find \emph{f} with certainty (see the third implication above). The result of FR is arrived at precisely because, even though each agent assumes a collapse happens when she measures, outside observers assume such measurements don't cause a collapse; that is, they assume H (for similar analyses see \cite{lazarovici,sudbery1,sudbery2}).

Is FR's result interesting? We don't think so. By depending on (H), the argument has a very limited range of applicability. There are a few interpretations, such a Relational Quantum Mechanics or Consistent Histories, that do seem to assume (H), or something similar, but they also try to negate either (C) or (S) explicitly. It is not clear if any of these approaches is really satisfactory, what is clear is that FR's result does not help clarify the issue. In sum, no known interpretation that is unambiguous regarding the collapse postulate is really affected by FR's result.

In any case, what we want to show next is that a result fully analogous to the one constructed in \cite{FR} can be arrived at via a much simpler arrangement, namely a simple EPR scenario (in Bohm's formulation in terms of spin-$\frac{1}{2}$ particles, \cite{Bohm}).

\section{The EPR-Wigner experiment}
\label{EPR}
Consider the following experimental procedure: 
\begin{itemize}
\item F prepares a singlet inside of her lab, then she sends to W one of the particles of the pair and measures the other.
\end{itemize}
That's it. Now, given that the state is a singlet, the following implication follows:
\begin{equation}\label{I5}
\text{If F finds $\uparrow$, she knows that W will find $\downarrow$ with certainty.}
\end{equation}
However, at the same time, given (H), F knows that, even after F measures, W assigns to the pair of particles a singlet state. From this it follows that, even if F finds $\uparrow$, F knows that W predicts a 50$\%$ chance to find $\uparrow$. As a result, by (C), even if F finds $\uparrow$, F predicts that W has a 50$\%$ chance to find $\uparrow$. But this, of course, is inconsistent with (\ref{I5}).

In sum, by considering this extremely simple scenario, and without having to consider measurements over complex, macroscopic systems, we arrive at a contradiction which is fully analogous to the one encountered by FR. After all, if we think about it, it comes as no surprise that if one considers an EPR pair, and one assumes, with (H), that the measurement on one particle does not change the state of the other, then inconsistencies may arise.

In order to remove any lingering doubts that our results is in fact fully equivalent to FR's, we can dress-up our setting to make it more like theirs. In particular, we could consider, as FR, two labs, four observers and a communication channel connecting the labs. Then we could have $\overline{\text{F}}$ measure a quantum coin prepared in the state $\frac{1}{\sqrt{2}}\left(\ket{h}+\ket{t} \right)$ and have her send to F an electron in the state $\ket{\downarrow}$, if she finds \emph{h}, and $\ket{\uparrow}$, if she finds \emph{t} (all this is equivalent to $\overline{\text{F}}$ preparing a singlet, sending away one of the particles and measuring the other). Then F would measure the spin of the electron in the $\{\ket{\downarrow},\ket{\uparrow}\}$ basis (this step is superfluous). Next $\overline{\text{W}}$ would measure the state of $\overline{\text{F}}$'s lab in the $\{\ket{h}_{\overline{\text{L}}},\ket{t}_{\overline{\text{L}}}\}$ basis and would announce her result (this allows for W to know $\overline{\text{F}}$'s result so that W can arrive at a contradiction, but this is not necessary as $\overline{\text{F}}$ himself can find a contradiction). Finally, W would measure the state of F's lab in the $\{\ket{\downarrow}_{\text{L}},\ket{\uparrow}_{\text{L}}\}$ basis. Then, by a reasoning fully analogous to that used by FR, we would conclude that if $\overline{\text{W}}$ announces that she found, say, \emph{t}, then W will predict that she will find $\uparrow$ with certainty. However, by (H), she will also compute the state of both labs to be a superposition and predict a non-zero chance to find $\downarrow$ (just as in the EPR case, $\overline{\text{F}}$ can reason analogously to arrive at a contradiction).

So what can we say about the fact that FR's result can be arrived at in a simple EPR scenario? Well, as is well-known (and as is clear from the title), the primary objective of the EPR paper was to assess whether standard quantum mechanics is complete. To do so, the authors (implicitly) assume a locality principle that could be summarized as follows:
\begin{description}[font=\normalfont,labelindent=.5cm]
\item[(L)] If two systems are far apart, then actions on one cannot have instantaneous effects on the other.
\end{description}
Note that, within an EPR scenario, with two entangled particles sent to two separate locations, by assuming (L), one concludes that an action on one of the particles cannot affect the other and, in particular, that a measurement on one of the particles cannot cause a collapse on the other. Clearly this coincides with what the assumption (H) would predict for such a scenario. That is, in an EPR scenario, to assume (L) leads to the same consequences as assuming (H). If this is so, we conclude that what one learns from FR's work must be equivalent to what one learns from the EPR paper. So, what does one learn from the EPR paper? As we said, it seeks to assess the completeness of standard quantum mechanics and concludes, implicitly assuming (C) and (S), that (L) and completeness are inconsistent. We see that, if we identify (L) and (H) in this scenario, then FR's result is indeed equivalent to that of the EPR paper.

It is important to point out that the recognition that FR's result is equivalent to what one can learn from an EPR scenario, implies that any theory that satisfactorily explains the latter will have no trouble regarding the former. That is, any theory with resources to predict consistent results for an EPR experiment will stay clear from trouble regarding FR's no-go result.

Before moving on, we would like to compare the disclosed equivalence between FR and EPR, with the claim in \cite{Drezet} that FR's work can be read as a rephrasing of Hardy's theorem \cite{Hardy1,Hardy2} (in fact, FR's proposal is said to build upon Hardy's work). We believe, however, that exactly in the same way that Bell's theorem is stronger than EPR, Hardy's result is stronger than FR's. As we explained above, EPR proves that locality and completeness are inconsistent, but leaves the door open for local hidden-variable theories to be viable. In \cite{Bell1964}, building on that, Bell demonstrates that all local theories are incompatible with the predictions of standard quantum mechanics. Hardy's theorem constitutes an alternative proof of Bell's theorem for the case of deterministic local theories, so, with Bell, goes further than EPR. FR's work, on the other hand, shows that completeness and (H) are inconsistent, but does not even attempt to prove that all theories satisfying (H) must make different predictions than standard quantum mechanics.\footnote{To prove  this, as Bell or Hardy, and unlike FR, one would need to introduce state descriptions other than those of standard quantum mechanics and would have to allow for alternative measurement settings for different runs of the experiment. It is also important to point out that Hardy's proof works for all entangled states, \emph{except the singlet}. Therefore, the fact that we constructed a scenario equivalent to that of FR, but with a singlet, shows that FR's result cannot be equivalent to Hardy's.} It seems clear, then, that FR is not equivalent to Hardy's theorem---the latter is strictly stronger.

\section{Conclusions}
\label{C}

Frauchiger and Renner try to show that ``quantum theory cannot consistently describe the use of itself''. To do so, they consider an elaborate, extended Wigner's friend scenario involving four agents, two labs and the supposition that quantum mechanics can be applied to macroscopic systems. Elsewhere it has been shown that FR's result is not as robust or general as intended, as it depends on a highly non-trivial assumption regarding the collapse mechanism---namely, that when a measurement is carried out inside of a closed lab, such a measurement leads to a collapse for inside observers, but does not do so for outside observers. 

FR's proposal trades on the vague nature of the collapse postulate---on the ambiguity as to what causes it and what it entails---to arrive at a conflict. That puts FR in the company of many ``results'' of the past hundred years, such as delayed choice arrangements or quantum erasers, that also exploit the conceptual lagoons of standard quantum mechanics to arrive at alleged contradictions or paradoxes. Of course, as soon as one resolves such conceptual lagoons in a satisfactory way, all those problematic results disappear.

In this work we have shown that FR's result can be derived in a much simpler setting, and without having to assume that observers and labs can enter into Schrödinger cat-like states. That is, we have shown that a result fully analogous to that of FR can be arrived at by considering a simple EPR scenario. Moreover, we have argued that any lesson that could be learned from FR's result is essentially contained in the original EPR paper. We conclude that FR's result is even less illuminating than previously recognized as it does not offer any novel insights into the conceptual problems of quantum theory.

\section*{Acknowledgments}

E.O. acknowledges support from UNAM-PAPIIT grant IN102219.


\bibliographystyle{plain}
\bibliography{bibFR.bib}


\end{document}